\documentclass[a4paper,10pt]{article}
\usepackage{spconf,amsmath,graphicx,cite}
\usepackage[german,english]{babel}
\usepackage{psfrag}
\usepackage{cite}
\usepackage{epsfig}
\usepackage{amssymb}
\usepackage{bbm}
\usepackage{algpseudocode}
\usepackage{array}

\usepackage{tikz}

\usetikzlibrary{arrows,shapes,positioning,
arrows,decorations.markings,calc}
\usetikzlibrary{dsp,chains}
\usetikzlibrary{arrows}

\usepackage{pgfplots}
\usepackage{standalone}

\newcommand{\vect}[1]{\mathbf{#1}}
\newcommand{\matt}[1]{\mathbf{#1}}

\newcommand{\jim}{\mathrm{j}\,}
\usepackage{mathtools}
\DeclarePairedDelimiter{\norm}{\lVert}{\rVert}

\DeclareMathOperator*{\argmin}{arg\,min}

\DeclareMathOperator*{\E}{E}
\newcommand{\T}{\operatorname{\mathrm{T}}}

\title{DFE/THP duality for FBMC with Highly Frequency Selective Channels}
\name{Hela Jedda, Leonardo G. Baltar, Oliver De Candido, Amine Mezghani, Josef A. Nossek\thanks{The authors acknowledge the financial support by the EU FP7-ICT project EMPhAtiC (http://www.ictemphatic.eu) under grant agreement no. 318362.}}
\address{Institute for Circuit Theory and Signal Processing\\
Technische Universit\"at M\"unchen\\
80290 Munich, Germany\\
Email: \{hela.jedda, leo.baltar\}@tum.de}

\begin{document}
\maketitle

\tikzset{DSP lines/.style={help lines,very thick,color=black}}
\tikzset{line_arrow/.style={help lines,very thick,color=black,->,-angle 90}}
\tikzset{filter/.style={rectangle,inner sep=0pt,minimum height=0.8cm,minimum width=1cm,draw=black,very thick, text centered}}
\tikzset{delay/.style={rectangle,inner sep=0pt,minimum size=0.8cm,draw=black,very thick}}
\tikzset{downsampling/.style={rectangle,inner sep=0pt,minimum size=0.8cm,draw=black,very thick}}
\tikzset{upsampling/.style={rectangle,inner sep=0pt,minimum size=0.8cm,draw=black,very thick}}
\tikzset{empty_node/.style={inner sep=0pt,minimum size=0cm}}
\tikzset{connection/.style={circle,draw=black,fill=black,inner sep=0pt,minimum size=2mm}}
\tikzset{coefficient/.style={isosceles triangle,draw=black,very thick,inner sep=0pt,minimum size=.5cm}}
\tikzset{source/.style={semicircle,minimum size=.5cm,draw=black,very thick,shape border rotate=270}}
\tikzset{adder/.style={circle,minimum size=.25cm,inner sep=0pt,draw=black,very thick}}
\tikzset{multiplier/.style={circle,minimum size=.25cm,inner sep=0pt,draw=black,very thick}}
\tikzset{double_arrow/.style={double distance=5pt,thick,shorten >= 6pt,decoration={markings,mark=at position 1 with {\arrow[scale=.6,>=angle 90]{>}}},postaction={decorate}}}
\selectlanguage{english}
\begin{abstract}
Filter bank based multicarrier with Offset-QAM systems (FBMC/OQAM) are strong candidates for the waveform of future 5-th generation (5G) wireless standards. These systems can achieve maximum spectral efficiency compared to other multicarrier schemes, particularly in highly frequency selective propagation conditions. In this case a multi-tap, fractionally spaced equalizer or precoder needs to be inserted in each subcarrier at the receiver or transmitter side to compensate inter-symbol interference (ISI) and inter-carrier interference (ICI). In this paper we propose a new Tomlinson-Harashima precoder (THP) design for FBMC/OQAM based on the mean squared error (MSE) duality from a minimum MSE (MMSE) designed decision feedback equalizer (DFE).
\end{abstract}
\section{Introduction}
\label{sec:intro}

In recent years multi-carrier systems have been at the forefront of communication systems due to their attractive properties at high data rates. Orthogonal frequency division multiplexing with a cyclic prefix (CP-OFDM) is a widely implemented solution for multi-carrier systems in standards such as IEEE 802.11, LTE or VDSL. Its popularity is partly due to the simple equalization enabled by the CP and the efficient implementation using Fast Fourier Transform (FFT). However, this comes at the price of a loss in spectral efficiency due to the CP, which is extremely long in the presence of highly frequency selective channels. CP-OFDM additionally suffers from high out-of-band emissions and the necessity of perfect synchronization.

An alternative solution to CP-OFDM are FBMC/OQAM systems which are a strong contender for 5G mobile communication systems \cite{6923528}. FBMC/OQAM systems have improved spectral efficiency due to the Synthesis and Analysis Filter Banks (SFB and AFB) at the transmitter and receiver \cite{Siohan2002}, which guarantees higher selectivity in the frequency domain and a much lower out-of-band radiation compared with CP-OFDM \cite{Baltar2007}. This form of pulse shaping limits the ICI, whilst simultaneously attributing to more ISI within each individual sub-carrier. Furthermore, FBMC/OQAM systems are extremely efficient in the presence of highly frequency selective channels. These advantages over CP-OFDM come at the cost of slightly higher computational complexity, however, this is not problematic \cite{Baltar2011}.

In \cite{Waldhauser2008a} an MMSE-based, multi-tap, per sub-carrier, linear equalizer design for a Single Input Single Output (SISO) setting was introduced to compensate the ISI and ICI in the presence of highly frequency selective channels. In \cite{Baltar2009a} a DFE extension of \cite{Waldhauser2008a} was proposed. In situations where the channel impulse response is known at the transmitter side, it is usually preferred to employ a precoder in order to avoid noise coloring and noise power amplification at the receiver. Additionally, if there is a symmetry in the computational power, e.g.~cellular networks, where most of the processing can be concentrated at the base station, i.e.~precoding in the downlink (DL) and equalization in the uplink (UL). In \cite{Caus2012} the authors proposed MIMO-FBMC precoders based on the signal-to-leakage plus noise ratio (SLNR) and on the signal-to-interference plus noise ratio
(SINR). In \cite{Horlin2013} three precoders are inserted in each subcarrier to combat ICI. In \cite{Caus2013} a THP is proposed for frequency flat channels in each subcarrier. In \cite{DeCandido2015} linear precoders based on the MSE duality with the linear equalizer was extended for the Multi-User Multiple Input Single Output (MU-MISO) setting. In this contribution we propose a THP design for SISO FBMC/OQAM systems using the MSE 
duality and the DFE design from \cite{Baltar2009a}.

This paper is organized as follows: in Section \ref{sec:System_Model} we introduce the FBMC system model. In Section \ref{sec:DFE} we calculate the MSE for the DFE and show its MMSE solution as in \cite{Baltar2009a}. In Section \ref{sec:THP} we explain the THP and define its MSE. We explore two different methods to transform the DFE into a dual THP in Section \ref{sec:Duality_transformation}. In Sections \ref{sec:simulation_results} and \ref{sec:Conclusions} we interpret the simulation results and summarize this work. 

Notation: Bold letters indicate vectors and matrices, non-bold letters express scalars. The operators $(.)^{*}$, $(.)^{\rm T}$, $(.)^{\rm H}$ and $\E\left[.\right]$ stand for complex conjugation, the transposition, Hermitian transposition and the expectation, respectively. The $n \times n$ identity matrix is denoted by $\matt{I}_{n}$ while the zeros (ones) matrix with $n$ rows and $m$ columns is defined as $\matt{0}_{n,m}$ ($\matt{1}_{n,m}$). We define $\left(\bullet\right)^{(R)} = \Re \lbrace \bullet \rbrace$, $\left(\bullet\right)^{(I)} = \Im \lbrace \bullet \rbrace$. The operator $\bar{\vect{x}}$ concatenates vertically the real- and imaginary valued parts of the vector/matrix $\vect{x}$. The vector $\vect{e}_{l}$ represents a zero vector with 1 in the $l$-th position. 

\section{FBMC System Model}
\label{sec:System_Model}
In a SISO FBMC system, the SFB combines the $M$ complex valued QAM input signals $d_k[m]$, $k=1,...,M$, generated at a rate of $1/T_\text{s}$, into a single complex valued signal $t[r]$ at a higher rate of $M/T_\text{s}$. The signal is transmitted across a highly frequency selective additive white Gaussian noise channel to the receiver. In our system, $M$ corresponds to the total number of sub-carriers and $M_{\text{u}}$ the number of sub-carriers we transmit across. The AFB separates the received signal back into its $M_\text{u}$ components at the lower rate $1/T_\text{s}$ per sub-carrier.

The first operation in the SFB is the O-QAM staggering of the input $d_k[m]$ and the output sequence $\vect{x}_k[n]$ reads as
{\small\begin{align*}
\vect{x}_k[n] = \begin{cases} 
\begin{bmatrix}
\! \alpha_k [m] \!\!&\!\! \mathrm{j} \beta_k[m] \!\!&\!\! \alpha_k [m-1] \!\!&\!\! \cdots \!
\end{bmatrix}^T, &\!\! \text{$k+n$ is odd}, \\
\begin{bmatrix}
\! \mathrm{j} \beta_k[m] \!\!&\!\! \alpha_k [m] \!\!&\!\! \mathrm{j}\beta_k [m-1] \!\!&\!\! \cdots \!
\end{bmatrix}^T, & \!\!\text{$k+n$ is even}.
\end{cases}
\end{align*}}
The input symbol  $d_k[m]$ is split into its real $d^{(R)}_k[m] = \alpha_k [m]$ and imaginary $\mathrm{j} d^{(I)}_k[m] = \mathrm{j} \beta_k[m]$  parts, up-sampled by a factor of 2, then depending on which sub-carrier we observe, either $\alpha_k [m]$ or $\mathrm{j} \beta_k[m]$ symbol is delayed by exactly $T_s/2$ and finally these components are added together. 
Due to this characteristic of the O-QAM symbols, there is a phase change of $\pi /2$ between immediately adjacent sub-carriers, ensuring orthogonality between them. At the receiver, the AFB applies O-QAM de-staggering to reconstruct the complex QAM $\hat{d}_k[m]$ symbols from the equalized $\hat{x}_k[n]$ symbols.


Since an implementation as described above is not very efficient due to the extremely high data rate, an implementation as a Modified DFT (MDFT) filter bank is much more efficient. An MDFT filter bank takes advantage of exponentially modulated, pulse shaping filters given by
{\small
\begin{align*}
h_k[r]\! =\! h_{\text{p}}[r] \; \text{exp} \; \left(\mathrm{j} \; \frac{2 \pi}{M}\;k\; \left(r - \frac{L_{\text{p}}-1}{2}\right)\right), \; r = 0,\ldots,L_{\text{p}}-1,
\end{align*}}
where $h_{\text{p}}[r]$ is a lowpass prototype filter with length $L_{\text{p}}=KM+1$, with $K$ representing the overlapping factor to indicate the number of symbols which overlap in time. $K$ should be kept as small as possible not only to limit the complexity but also to reduce the time-domain spreading of the symbols and the transmission latency. Furthermore, MDFT takes advantage of the polyphase decomposition of $h_{\text{p}}[r]$ so that the filtering can be performed at a rate of only $2/T_s$. 

To minimize the complexity in the calculations of the equalizer and precoder vectors, we set $K = 4$ and the roll-off factor of our root raised cosine filter equal to one.  Thus, the frequency response of the filter $h_k[r]$ only significantly overlaps with the two adjacent filters.

For a simple total transmission notation we define the following filtering and downsampling operation, $h_{l,k}[n] \!=\! \left( h_l\!\ast\! h_{\text{ch}}\!\ast\! h_k \right)[r]\mid_{r=n \frac{M}{2}}$. This represents the total impulse response from the sub-carrier $l$ at the transmitter into the sub-carrier $k$ at the receiver, with $l \in \lbrace k\!-\!1, k, k\!+\!1 \rbrace$. The resulting filter has length $N=\left\lceil \frac{2(L_{\text{p}} - 1)+L_{\text{ch}}}{M/2} \right\rceil$.

In the following sections we work with a purely real notation and therefore define a purely real input sequence as $\vect{x}'_k[n]$ where the relation $\vect{x}_k [n] = \matt{J}_{k,n} \vect{x}'_k[n]$ holds with 
\begin{align*}
\matt{J}_{k,n} = \begin{cases}
\text{diag} 
\begin{bmatrix} 1 & \jim & 1 & \jim &\cdots \end{bmatrix} , 
& \text{$k+n$ is odd}, \\
\text{diag} 
\begin{bmatrix} \jim & 1 & \jim & 1 & \cdots \end{bmatrix} , & \text{$k+n$ is even}.
\end{cases}
\end{align*}
This extracts the imaginary $\mathrm{j}$'s from the input signal. We then multiply the transposed convolution matrix of $h_{l,k}[n]$ by $\matt{J}_{k,n}$ and are left with $\matt{H}'_{l,k} =  \matt{H}_{l,k} \matt{J}_{k,n}$.


\section{Decision Feedback Equalizer}
\label{sec:DFE}
We are interested in designing a multi-tap decision feedback equalizer per sub-carrier with feed-forward (FF) filter $\vect{f}^{\text{UL}}_{2,k}$ of length $L_{\text{f}}$ and feedback (FB) filter $\vect{b}^{\text{UL}}_{k}$ of length $L_{\text{b}}$ and a single-tap precoder $f^{\text{UL}}_{1,k}  \in \mathbb{R}_+$ per sub-carrier. We assume there is no channel state information at the transmitter and, therefore, set the scalar precoder to one, i.e.~ $f^{\text{UL}}_{1, k}= 1$. The received and equalized signal in sub-carrier $k$, when $n+k$ is odd, i.e.~for $\hat \alpha_k[m]$, is defined as
\begin{align}\label{eq:DFE_output}
\hat{x}_k[n] =\vect{w}^{\rm T}_k\left(\matt{A}_k\vect{x}_{1,k} + \matt{B}_k\vect{x}_{2,k} + \matt{\Xi}_k\bar{\boldsymbol{\eta}}\right),
\end{align}
where we define
\begin{align}
\vect{w}_k &= \begin{bmatrix} \bar{\vect{f}}^{\text{UL}}_{2,k}  \\ (\vect{b}^{\text{UL}}_{k})^{(R)} \end{bmatrix} \in \mathbb{R}^{(2 L_{\text{f}}+L_{\text{b}})} , \nonumber \\
\matt{A}_k &= \begin{bmatrix}
			\bar{\matt{H}}'_{k,k} \!\!\!&\!\!\! \matt{0}_{2L_f \times L_b} \\
			\matt{0}_{L_b \times (N+L_f-1)} \!\!\!&\!\!\! -\matt{I}_{L_{\text{b}}}
			\end{bmatrix} \!\!\in\! \mathbb{R}^{(2 L_{\text{f}}+L_{\text{b}}) \!\times\! (N+L_{\text{f}}+L_{\text{b}}-1)},\nonumber \\
\matt{B}_k &= \begin{bmatrix}
			\bar{\matt{H}}'_{k-1,k}  \!\!\!&\!\!\!\bar{\matt{H}}'_{k+1,k}\\
			\matt{0}_{L_b \times 2(N+L_f-1)}
			\end{bmatrix}\in \mathbb{R}^{(2 L_{\text{f}}+L_{\text{b}}) \times 2(N+L_{\text{f}}-1)}, \nonumber \\
\matt{\Xi}_k &=\begin{bmatrix}
			\matt{\Gamma}_k\\
			\matt{0}_{L_b \times 2(L_f+L_p-1)}
			\end{bmatrix}, \nonumber \\
\matt{\Gamma}_k &= \begin{bmatrix}
			(\matt{H}_{k})^{(R)} &  -(\matt{H}_{k})^{(I)}\\
			(\matt{H}_{k})^{(I)} &  (\matt{H}_{k})^{(R)}
			\end{bmatrix} \in \mathbb{R}^{2 L_{\text{f}} \times 2(L_{\text{f}}+L_p\!-\!1)},\nonumber\\
\vect{x}_{1,k} &=  \begin{bmatrix}  \vect{x}'_k[n] \\ \vect{x}'_k [n- (\nu+1)] \end{bmatrix}\in \mathbb{R}^{(N+ L_{\text{f}}+L_{\text{b}}-1)},  \nonumber\\
\vect{x}_{2,k} &=  \begin{bmatrix} \vect{x}'_{k-1}[n] \\ \vect{x}'_{k+1}[n] \end{bmatrix}\in \mathbb{R}^{2(N+ L_{\text{f}}-1)},
\end{align}
where $\matt{H}_{k}$ is the $M/2$ downsampled, transposed convolution matrix of $h_k[r]$ which filters the noise $\boldsymbol{\eta}$ and $\nu$ is the equalization latency in our system. For the following derivations we assume the input signals to be independent and identically distributed (i.i.d.) and Gaussian distributed. The covariance matrix of $\vect{x}'_k [n]$ is defined as $\E \left[ \vect{x}'_k [n] \vect{x}'^{\rm T}_k [n] \right]= \sigma^2_x \matt{I}$. Furthermore, we assume the additive noise is Gaussian distributed with $\boldsymbol{\eta} [n] \sim \mathcal{N}_{\mathbb{C}} \left(\matt{0},\sigma^2_{\eta} \matt{I}\right)$.

The MMSE equalizer is given by
\begin{align}
\vect{w}_k &= \argmin_{\vect{w}_k} \epsilon^{\text{UL}}_{k} \nonumber \\
&= \argmin_{\vect{w}_k} \E\left\{ \Vert \hat{x}_k[n] -x_k[n-\nu] \Vert^2_2\right\} \nonumber \\
&=\!\! \left(\matt{A}_k \matt{\Psi}\matt{A}^{\rm T}_k + \sigma^2_x \matt{B}_k\matt{B}^{\rm T}_k + \frac{\sigma^2_{\eta}}{2} \matt{\Xi}_k\boldsymbol{\Xi}^{\rm T}_k\right)^{-1} \!\! \matt{A}_k \matt{\Psi} \boldsymbol{e}_{\nu+1}, \nonumber
\end{align}
where
\begin{align}
&\matt{\Psi} = \E\left[\Vert \vect{x}_{1,k}\vect{x}^{\rm T}_{1,k} \Vert^2_2\right] = \sigma^2_x \begin{bmatrix} \matt{I}_{N+L_{\text{f}}-1} & \matt{\Upsilon} \\
																				 \matt{\Upsilon}^{\rm T} & \matt{I}_{L_{\text{b}}}
												 \end{bmatrix} \nonumber \\
&\matt{\Upsilon} = \begin{cases}
\left[ \begin{array}{ccc} \!\!\matt{0}_{L_{\text{b}}\times (\nu+1)} \!\!&\!\! \matt{I}_{L_{\text{b}}} \!\!&\!\! \matt{0}_{L_{\text{b}}\times (L -L_{\text{b}})} \end{array} \!\!\right]^{\rm T}\!\!\!\!&\!\! \text{, if } L>L_{\text{b}} \\
\left[ \begin{matrix} \matt{0}_{L_{\text{b}} \times (\nu+1)} & \matt{I}_{L_{\text{b}}}\end{matrix}\right]^{\rm T}   \!\!\!\!&\!\! \text{, if } L=L_{\text{b}} \\
\begin{bmatrix} \matt{0}_{(\nu+1)\times L_{\text{b}}} \\
																				\matt{I}_{L}				&													\matt{0}_{L\times (L_{\text{b}}-L)} \end{bmatrix} \!\! &\!\! \text{, if } L<L_{\text{b}},
									 \end{cases}
\end{align}
where $L=L_{\text{f}}+N-\nu-2$.
The MSE can be expressed by
\begin{align}
\epsilon^{\text{UL}}_{k} =& \:\bar{\vect{f}}^{\text{UL},\rm T}_{2,k} \left(\sigma^2_x \sum^{k+1}_{l=k-1} \bar{\matt{H}}'_{l,k} \bar{\matt{H}}'^{\rm T}_{l,k} +\frac{\sigma^2_{\eta}}{2} \matt{\Gamma}_k \matt{\Gamma}_k^{\rm T} \right) \bar{\vect{f}}^{\text{UL}}_{2,k} \nonumber\\
&+ \sigma^2_x \left( \vect{r}^{\rm T} \vect{r} - 2  \left(  \bar{\vect{f}}^{\text{UL},\rm T}_{2,k} \bar{\matt{H}}'_{k,k} \vect{r}  \right)\right), \nonumber \\
\text{with } & \vect{r}^{\T} = \begin{bmatrix} \vect{0}_{1 \times \nu} & 1 & (({\vect{b}^{\text{UL}}_k})^{\T})^{(R)}& \vect{0}_{1 \times (L-L_b)} \end{bmatrix}.
\end{align}

It can be shown, that calculating the precoder or equalizer filters with either the real part $\hat \alpha_k[m]$ ($n+k$ is odd) or imaginary part $\hat \beta_k[m]$ ($n+k$ is even) result in the same solution \cite{Waldhauser2008a,Newinger2014a}.

\section{Thomlinson Harashima Precoder}
\label{sec:THP}
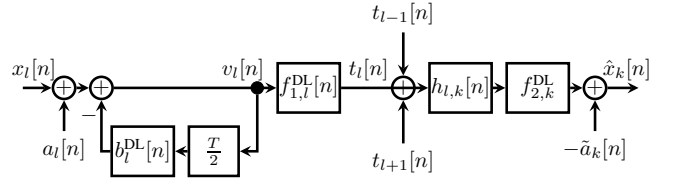
\begin{figure}[t]
\centering
\resizebox{\columnwidth}{!} {%
\begin{tikzpicture}
\node (in1){};
\node  (input1) [right=of in1] [xshift=-1.3cm, yshift=0.3cm]{$x_l[n]$};
\node [adder] (add1)  [xshift=-0.2cm][right=of input1][xshift=-1cm,yshift=-0.3cm]{$+$};
\node (mod1)[below=of add1][yshift=0.5cm]{$a_l[n]$};
\node [right=of add1][xshift=-1.2cm, yshift=0.3cm]{};
\node [adder] (add) [right=of add1][xshift=-0.8cm]{$+$};
\node  [below=of add][xshift=-0.2cm, yshift=1cm]{$-$};

\node [connection](out1) [right=of add][xshift=1.2cm]{};
\node [right=of out1][xshift=-1.8cm,yshift=0.3cm]{$v_l[n]$};
\node [delay] (delay) [below=of out1][xshift=-0.7cm,yshift=0.5cm] {$\frac{T}{2}$};
\node [filter] (fb)[left=of delay] [xshift=0.8cm]{${b}^{\text{DL}}_l[n]$};
\node [filter] (ff1)[right=of out1][xshift=-0.8cm]{${f}^{\text{DL}}_{1,l}[n]$};
\node (t) [right=of ff1][xshift=-1cm,yshift=0.3cm]{$t_l[n]$};
\node [adder] (add4) [right=of ff1][xshift=-0.2cm]{$+$};
\node (sc1)[above=of add4][yshift=-0.3cm]{$t_{l-1}[n]$};
\node (sc3)[below=of add4][yshift=0.3cm]{$t_{l+1}[n]$};
\node [filter] (ch) [right=of add4] [xshift=-0.8cm]{$h_{l,k}[n]$};
\node [filter] (ff) [right=of ch][xshift=-0.8cm]{$f^{\text{DL}}_{2,k}$};
\node [right=of ff][xshift=-1cm, yshift=0.3cm]{};
\node [adder] (add2) [xshift=-0.8cm][right=of ff]{$+$};
\node (mod2)[below=of add2] [yshift=0.5cm]{$-\tilde{a}_k[n]$};
\node (output1) [right=of add2][xshift=-1.2cm,yshift=0.3cm]{$\hat{x}_k[n]$};
\node (output)[right=of add2][xshift=-0.5cm]{};
\draw[DSP lines] [-stealth] (in1.east) -- (add1.west);
\draw[DSP lines] [-stealth] (add1.east) -- (add.west);
\draw[DSP lines] [-stealth] (add.east) --  (ff1.west);
\draw[DSP lines] [-stealth] (ff1.east) -- (ch.west);
\draw[DSP lines] [-stealth] (ch.east) -- (ff.west);
\draw[DSP lines] [-stealth] (ff.east) -- (add2.west);
\draw[DSP lines] [-stealth] (add2.east) -- (output.west);
\draw[DSP lines] [-stealth] (out1.south) |- (delay.east);
\draw[DSP lines] [-stealth] (delay.west) -- (fb.east);
\draw[DSP lines] [-stealth] (fb.west) -| (add.south);
\draw[DSP lines] [-stealth] (mod1.north) -- (add1.south);
\draw[DSP lines] [-stealth] (mod2.north) -- (add2.south);
\draw[DSP lines] [-stealth] (sc1) -- (add4.north);
\draw[DSP lines] [-stealth] (sc3) -- (add4.south);
\end{tikzpicture}%
} 
\caption{FBMC with THP subcarrier model}
\label{fig:thp}
\end{figure}

We define a per sub-carrier, multi-tap THP with FF filter $\vect{f}^{\text{DL}}_{1,k}$ and FB filter $\vect{b}^{\text{DL}}_{k}$ and a per sub-carrier single-tap equalizer $f^{\text{DL}}_{2,k}  \in \mathbb{R}_+$ at the receiver. Since we have a feedback loop in the THP, we are at risk of stability problems. Therefore, we additionally introduce the modulo operator $M(.)$ at the transmitter and the receiver in the DL scenario, that upper bounds the output signals.  The modulo operator that is adapted to the O-QAM structure is expressed as
\begin{equation*}
M(x_l[n]) \!=\! \begin{cases} x^{(R)}_{l}[n] - \left\lfloor \frac{x^{(R)}_{l}[n]}{\tau}+\frac{1}{2} \right\rfloor \tau, & \text{if $l+n$ is odd }\\
jx^{(I)}_{l}[n] - j\left\lfloor \frac{x^{(I)}_{l}[n]}{\tau}+\frac{1}{2}\right\rfloor \tau,  & \text{if $l+n$ is even},
\end{cases}
\end{equation*}\\
where $\tau \in \mathbb{R}_+$ is the modulo constant and depends on the modulation alphabet \cite{Fischer}. The modulo operators can be replaced by the summations of the signals $a_l[n]$ and $-\tilde{a}_k[n]$, which are multiples of $\tau$, as shown in Fig. \ref{fig:thp}. The output of the modulo operator at the transmitter in sub-carrier $l$ is denoted by $v_l[n]$ and we define $\vect{v}_l[n] = \matt{J}_{l,n} \vect{v}'_l[n]$ with covariance matrix $\E\left[\vect{v}'_l[n] \vect{v}'^{\rm T}_l[n] \right] = \sigma^2_v \matt{I}$ and $\sigma^2_v = \frac{\tau^2}{12}$ \cite{Joham}. 

The MSE of the DL scenario can be expressed by
\begin{align}
\epsilon^{\text{DL}}_k =& \E\left\{\norm{\hat{x}_k[n] - x_k[n-\nu]}^2_2\right\} \nonumber \\
=& \: (f^{\text{DL}}_{2,k})^2\left(\sigma^2_v \sum^{k+1}_{l=k-1} \bar{\vect{f}}^{\text{DL},\rm T}_{1,l} \bar{\matt{H}}'_{l,k} \bar{\matt{H}}'^{\rm T}_{l,k} \bar{\vect{f}}^{\text{DL}}_{1,l} + \frac{\sigma^2_{\eta}}{2} \Vert\vect{h}_p\Vert^2_2\right) \nonumber \\
& + \sigma^2_v \left( \vect{s}^{\rm T}\vect{s} - 2  \left(f^{\text{DL}}_{2,k} \bar{\vect{f}}^{\text{DL},\rm T}_{1,k} \bar{\matt{H}}'_{k,k} \vect{s}   \right)\right)  \nonumber \\
\text{with } & \vect{s}^{\T} = \begin{bmatrix} \vect{0}_{1 \times \nu} & 1 & (({\vect{b}^{\text{DL}}_k})^{\T})^{(R)}& \vect{0}_{1 \times (L-L_b)} \end{bmatrix}.
\end{align}

\section{DFE/THP MSE duality Transformation}
\label{sec:Duality_transformation}
In this Section we investigate two different methods of transforming the DFE into an equivalent THP using the duality principle as introduced in \cite{Mezghani2006thp}. The basic idea behind an MSE duality transformation is to switch the roles of the UL and DL filters, i.e. we interchange each receiver filter in the UL scenario with the respective transmitter filter in the DL scenario. As the dual DL scenario has purely transmitter processing, we must ensure that the transmit power is subsequently limited, thus we weigh every transmitter filter with a strictly real constant and multiply the receiver with its inverse. 

These two duality transformation methods are summarized as follows:
\begin{itemize}
\item In Subsection \ref{sec:sum_MSE} we aim at conserving the Sum-MSE. This is the simplest form of duality since the Sum-MSE of all sub-carriers is kept equal when transforming the UL to a DL scenario. To this end, a single scaling factor is required which leads to relatively low computational complexity.
\item In Subsection \ref{sec:SC_MSE} we aim at conserving the Sub-Carrier MSE. In this method the MSE per sub-carrier is preserved when transforming the UL to a DL scenario resulting in an individual scaling factor for each sub-carrier. We get a linear system of equations for $M_{\text{u}}$ scaling factors, which results in a higher computational complexity than a Sum-MSE transformation.
\end{itemize}
In both of these MSE duality transformations the total transmit power is preserved, i.e. $ \sum_{k=1}^{M_{\text{u}}} \left\Vert \vect{f}^{\text{DL}}_{1,k} \right\Vert^{2}_2 \leq M_\text{u} $.

\subsection{Sum-MSE}
\label{sec:sum_MSE}
First, we define a relation between the UL and DL filters with a real-valued scaling factor for all sub-carriers such that
\begin{align*}
\bar{\vect{f}}^{\text{DL}}_{1,k} = \gamma \bar{\vect{f}}^{\text{UL}}_{2,k} \: ; \:
f^{\text{DL}}_{2,k} = \gamma^{-1} f^{\text{UL}}_{1,k} = \gamma^{-1} \: ; \: \vect{b}^{\text{DL}}_{k} = \vect{b}^{\text{UL}}_{k}.
\end{align*}
with $\gamma \in \mathbb{R}_+$ and recalling that the UL precoder scalar is set such that $f^{\text{UL}}_{1,k} = 1 ,\; \forall k$. To perform the Sum-MSE duality, the Sum-MSE is set equal between the UL and the DL scenario, i.e. we sum over all sub-carriers and set them equal
\begin{align} \label{eq:MSE_UL_DL_alpha_s}
\sum_{k=1}^{M_{\text{u}}}\epsilon^{\text{DL}}_k \stackrel{!}{=} \sum_{k=1}^{M_{\text{u}}}\epsilon^{\text{UL}}_k.
\end{align}
We get the following expression of the scaling factor $\gamma$
\begin{align}
\gamma^{2}   &= \frac{ M_{\text{u}} \frac{\sigma^2_{\eta}}{2}\Vert\vect{h}_\text{p}\Vert^2_2 }{\delta}, \text{ with}\nonumber \\
\delta &= \sum_{k=1}^{M_{\text{u}}} \bar{\vect{f}}^{\text{UL},\rm T}_{2,k} \left(\sigma^2_x \sum^{k+1}_{l=k-1} \bar{\matt{H}}'_{l,k} \bar{\matt{H}}'^{\rm T}_{l,k} +\frac{\sigma^2_{\eta}}{2} \matt{\Gamma}_k \matt{\Gamma}^{\rm T}_{k} \right)\bar{\vect{f}}^{\text{UL}}_{2,k} \nonumber \\
&- \left(\sigma^2_v \sum^{k+1}_{l=k-1} \bar{\vect{f}}^{\text{UL},\rm T}_{2,l} \bar{\matt{H}}'_{l,k} \bar{\matt{H}}'^{\rm T}_{l,k} \bar{\vect{f}}^{\text{UL}}_{2,l}  \right) \nonumber \\
&+ (\sigma^2_x -\sigma^2_v) \Bigg(\vect{r}^{\rm T} \vect{r} -2 \left( \bar{\vect{f}}^{\text{UL},\rm T}_{2,k} \bar{\matt{H}}'_{k,k} \vect{r}\right) \Bigg).
\end{align}

This form of duality transformation guarantees equality in the Sum-MSE between the UL and the DL scenario. Therefore, this method can be interpreted as allocating an equal amount of transmit power whilst spreading this transmit power across the sub-carriers as required.  The disadvantage of this method arises if the MSE of certain sub-carriers is disproportionately large. This leads to these sub-carriers obtaining a greater amount of transmit power.

\subsection{Sub-Carrier MSE (SC-MSE)}
\label{sec:SC_MSE}
Second, we define a relation between the UL and DL filters with a real-valued scaling factor per sub-carrier such that
\begin{align*}
\bar{\vect{f}}^{\text{DL}}_{1,k} &= \gamma_k \bar{\vect{f}}^{\text{UL}}_{2,k}  \: ; \:
f^{\text{DL}}_{2,k} &= \gamma^{-1}_{k} f^{\text{UL}}_{1,k} = \gamma^{-1}_{k}  \: ; \:
\vect{b}^{\text{DL}}_{k} &= \vect{b}^{\text{UL}}_{k},
\end{align*}
with $\gamma_k \in \mathbb{R}_+$ and recalling that the UL precoder is set such that $f^{\text{UL}}_{1,k} = 1 ,\; \forall k$.  We then set sub-carrier MSE equal between the UL and the DL scenario, i.e. we set the individual MSE expressions per sub-carrier equal such that
\begin{align}\label{eq:MSE_UL_DL_alpha_sk}
\epsilon^{\text{DL}}_k \stackrel{!}{=} \epsilon^{\text{UL}}_k, \quad \forall k.
\end{align}
We end up with a system of linear equations to solve for $M_{\text{u}}$ scaling factors $\gamma_k$
\begin{align}
\vect{\tilde{T}}\begin{bmatrix} \gamma^{2}_{1} & \cdots & \gamma^{2}_{M_{\text{u}}}\end{bmatrix}^{\T} = \frac{\sigma^2_{\eta}}{2} \left\|\vect{h}_\text{p}\right\|^2_2 \matt{1}_{M_\text{u}}, \text{ where}
\end{align}
{\footnotesize
\begin{align*}
\vect{\tilde{T}}_{k,j} \!\!=\!\!  \begin{cases}\left(\sigma^2_x-\sigma^2_v\right) \Big(\bar{\vect{f}}^{\text{UL},\rm T}_{2,k} \bar{\matt{H}}'_{k,k} \bar{\matt{H}}'^{\rm T}_{k,k} \bar{\vect{f}}^{\text{UL}}_{2,k} \vect{r}^{\rm T} \vect{r} \\
- 2  \bar{\vect{f}}^{\text{UL},\rm T}_{2,k} \bar{\matt{H}}'_{k,k} \vect{r}   \Big) + \bar{\vect{f}}^{\text{UL},\rm T}_{2,k}  \Big(\sigma^2_x \\
 \sum^{k+1}_{l=k-1, l\neq k} \bar{\matt{H}}'_{l,k} \bar{\matt{H}}'^{\rm T}_{l,k} +\frac{\sigma^2_{\eta}}{2} \matt{\Gamma}_k \matt{\Gamma}^{\rm T}_{k} \Big) \bar{\vect{f}}^{\text{UL}}_{2,k}   & \text{, if $j=k$}\\
\\ 
 - \sigma^2_v \bar{\vect{f}}^{\text{UL},\rm T}_{2,k-1} \bar{\matt{H}}'_{k-1,k} \bar{\matt{H}}'^{\rm T}_{k-1,k}  \bar{\vect{f}}^{\text{UL}}_{2,k-1}  & \text{, if $j=k-1$} \\
\\
- \sigma^2_v  \bar{\vect{f}}^{\text{UL},\rm T}_{2,k+1} \bar{\matt{H}}'_{k+1,k} \bar{\matt{H}}'^{\rm T}_{k+1,k}  \bar{\vect{f}}^{\text{UL}}_{2,k+1}  & \text{, if $j=k+1$}\\
0  & \text{else.}
 \end{cases}
\end{align*}}

This duality transformation guarantees that each SC-MSE remains equal for both UL and DL scenario. Therefore, we cannot spread the transmit power amongst the sub-carriers but instead we normalize the filter in each sub-carrier individually.

\section{Simulation Results}
\label{sec:simulation_results}
Throughout our simulations we  transmitted data across $M_{\text{u}} = 210$ of the available $M=256$ sub-carriers. The used multipath fading channel is based on bad urban area model (BU) with 6 taps. We used a sampling rate of $f_s = 15.36$MHz and a channel impulse response of duration $L_{\text{ch}} = 110$ samples. Thus, we had a subcarrier distance of $60$kHz. We used randomly generated $16$-QAM symbols and took a block length of $1000$ symbols per sub-carrier. For the used modulation scheme, we have following modulo constant value $\tau=8/ \sqrt{10}$. With the chosen system configurations, especially due to $L_{\text{ch}} = 110$ and the highly frequency selective channel, a CP-OFDM system would have required a CP with a minimum length of 109. This would have limited the data-throughput of the CP-OFDM to almost 50\%, therefore we have not included a direct comparison in the simulation results. Throughout the simulations we took the quantity of $E_b/N_0$ to be a pseudo-signal-to-noise ratio. We took the \
Bit Error Rate (BER) and MSE as an average 
over all sub-carriers. We took an average over $200$ randomly generated channel realizations.

We investigated six equalizers and precoders designs: a linear equalizer of length $L_{\text{lin}} = 9$ taps, its two dual precoders (SC-MSE and Sum-MSE), DFE with FF filter with $L_{\text{f}} = 7$ and FB filter with $L_{\text{b}} = 4$ taps and its two dual THPs. Note that the complexity of the FB filter is proportional to half of its length $L_{\text{b}}$, since its input and output are either purely real or purely imaginary.

\begin{figure}[t]
	\centering
    		\epsfig{file = 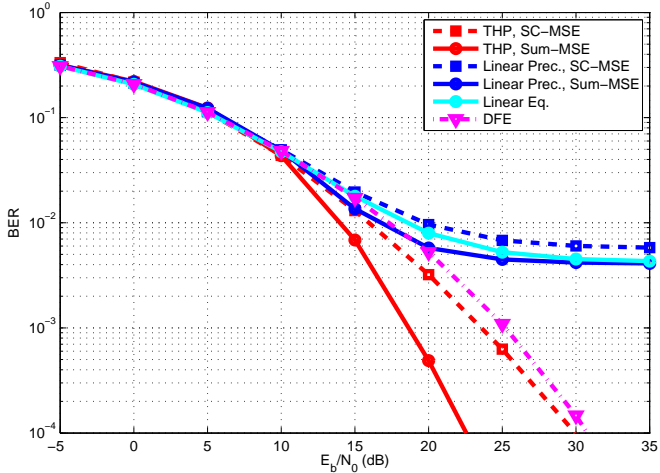, width = \columnwidth}
    		\caption{BER performance comparison}
	\label{fig:BER}
\end{figure}

\begin{figure}[t]
	\centering
    		\epsfig{file = 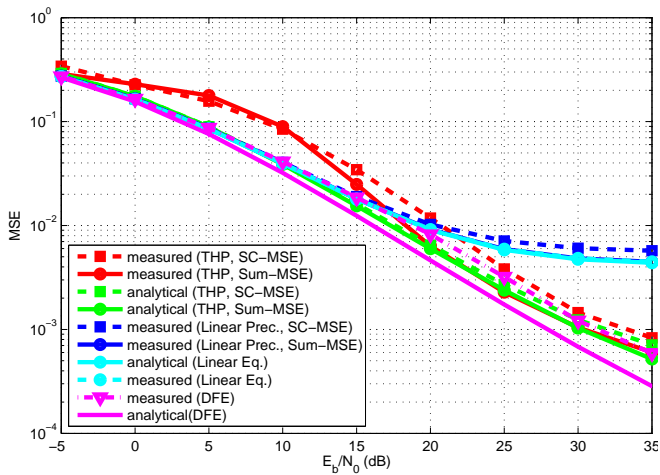, width = \columnwidth}
    		\caption{MSE performance comparison}
	\label{fig:MSE}
\end{figure}

As can be concluded from Fig.~\ref{fig:BER} and Fig.~\ref{fig:MSE}, the non-linear processing (DFE/THP) outperform the linear processing significantly in medium to high $E_b/N_0$ regime for the same complexity. Second, the Sum-MSE duality seems to be more beneficial than the SC-MSE duality. Third, the Sum-MSE dual THP compared to DFE shows an improved BER performance and similar behavior in terms of MSE.

\section{Conclusions}
\label{sec:Conclusions}
In this work, we proposed a new method to design a dual THP scheme from the DFE of a SISO FBMC system. To this end, we made use of the MSE duality transformation between an UL and a DL scenario, where either the Sum-MSE or the SC-MSE is conserved.  Throughout our simulations we observed first that the non-linear processing (DFE/THP) outperform the linear processing in medium to high $E_b/N_0$ region for the same complexity. Second, we have seen that the Sum-MSE duality performed the best in terms of BER over the whole $E_b/N_0$ regime, which could be explained by its equivalence to an inverse water-filling technique, i.e.~the subcarriers with poorer channel response are allocated more power.

\bibliographystyle{IEEEbib}
\bibliography{EUSIPCO_2015}

\end{document}